\documentstyle[12pt,epsf]{article}
\def\fun#1#2{\lower3.6pt\vbox{\baselineskip0pt\lineskip.9pt
\ialign{$\mathsurround=0pt#1\hfil##\hfil$\crcr#2\crcr\sim\crcr}}}

\newcommand{\be}{\begin{equation}}
\newcommand{\ee}{\end{equation}}
\newcommand{\bd}{\begin{displaymath}}
\newcommand{\ed}{\end{displaymath}}
\newcommand{\ba}{\begin{array}}
\newcommand{\ea}{\end{array}}
\newcommand{\bt}{\begin{tabular}}
\newcommand{\et}{\end{tabular}}

\begin{document}

\hfill{ITEP-TH-6/97}

\hfill TPI-MINN-97/04

\hfill NUC-MINN-97/2-T

\vspace{1.5cm}

\begin{center}
\large
QUARK CONDENSATE IN A MAGNETIC FIELD.
\end{center}

\begin{center} 
{\bf I.A. Shushpanov}

{\it Institute for Theoretical and Experimental Physics, 
B. Cheremushkinskaya 25, Moscow 117259, Russia}

\vspace{0.5cm}

and

\vspace{0.5cm}

{\bf A.V. Smilga}

\it{Institute for Theoretical and Experimental Physics, 
B. Cheremushkinskaya 25, Moscow 117259, Russia}

{\it and}

{\it TPI, School of Physics and Astronomy, University of Minnesota, MN
55455, USA}

\end{center}

\vspace{1.cm}
\centerline{February 1997}
\vspace{1.5cm}

\begin{abstract}
 We study the dependence of quark condensate $\Sigma$ on an external magnetic 
field. For weak fields, it rises linearly:
\be
\Sigma(H)=\Sigma(0)\left[1+\frac{eH \ln2}{16\pi^2 F_{\pi }^2}+
O\left(\frac{e^2H^2}{F_{\pi}^4}\right)\right]
\ee
$M_\pi$ and $F_\pi$ are also shifted so that the Gell-Mann -- Oakes -- Renner
relation is satisfied.

In the strong field region, $\Sigma(H)\propto (eH)^{3/2}$. 
\end{abstract}

\section{Introduction.}
 Phase structure of QCD is now a subject of an intense discussion. 
The question of how the properties of the system are modified by 
non-zero temperature was studied especially well (see e.g.
 \cite{review} for a 
recent review). We know now that, in the theory with massless quarks,
phase transition with restoration of spontaneously broken chiral 
symmetry occurs at some temperature $T=T_c$. That means that the order
parameter of the symmetry breaking, the quark condensate 
$\Sigma=-<\bar {q} q>$ falls down as temperature increases and turns 
to zero at $T\ge T_c$.
 When temperature is small compared to characteristic hadron scale 
$\mu _{hadr}$, the dependence $\Sigma(T)$ is known exactly 
\cite{condT}. In the
case of two massless flavors, 
\be
\Sigma(T)=\Sigma(0)\left[1-\frac{T^2}{8 F_{\pi}^2} 
-\frac{T^4}{384 F_{\pi}^4}- ...\right]
\label{conT}
\ee

The derivation of this formula relies on the fact that a lukewarm heat
bath involves mainly pions --- other degrees of freedom are not excited 
yet.  The pion interaction at small energies is known from the effective
chiral Lagrangian:
\be
L=\frac{F_{\pi}^2}{4} {\rm Tr} \{\partial_\mu U\partial_\nu U^\dagger\}+
\Sigma {\rm Re\ Tr} \{{\cal M}U^\dagger\} + \ {\rm higher \ order \ terms}
\label{CLLO}
\ee

Here U is a unitary SU(2) matrix ( it may be parameterized as 
$U= \exp\{i\tau^a \phi^a/F_\pi \}$ where $\phi^a$ is the pion field) and
we included also the mass term involving quark mass matrix $\cal M$. 
$F_\pi=93$ MeV is the pion decay constant and the parameter $\Sigma$
has the meaning of quark condensate $\Sigma=|<\bar{u}u>|=|<\bar{d}d>|$.
 The particular formula ($\ref{CLLO}$) 
of the Lagrangian is dictated by chiral 
symmetry (see \cite{Trento} for a nice pedagogical review).

 But temperature is not the only external parameter which can affect
the properties of the system. One can consider equally well a cold but 
dense system with non-zero mean  baryon charge density related to the
chemical potential. This system is less studied, but there are good
reasons to believe that also in this case chiral symmetry is restored
when the chemical potential exceeds some critical value.

In this paper, we address the issue of how the properties of QCD 
vacuum state depend on an external magnetic field. A naive expectation
based on the analogy with superconductivity and with the situation in 
QCD at non-zero temperature and/or chemical potential could be that
the condensate decreases as the magnetic field increases and melts down
completely at some critical value $H=H_c$ above which the chiral symmetry
is restored. We will see (and that is our main conclusion) that it is
 not so. Quark condensate  rises with the increase of magnetic field and no 
phase transition 
occurs.

The behavior of a hadron system in magnetic field was studied earlier by
Klevansky and Lemmer in the framework of Nambu-Jona-Lasinio (NJL) 
model \cite{Klev}. Solving gap equation in the presence of an external
 magnetic field
they found that the order parameters of the spontaneously broken chiral
symmetry, the dynamical fermion mass and the chiral condensate, rise
with the field. This conclusion was confirmed in later studies of NJL
model and related theories \cite{brat}. The shift of fermion mass and of the 
condensate is quadratic in field 
\be
\Sigma (H) = \Sigma(0)\left[1+c\frac{e^2H^2}{\Sigma^4}+
o\left(\frac{e^2H^2}{\Sigma^4}\right)\right]
\ee

So, in this model, chiral symmetry is not restored by magnetic field.
 Qualitatively, NJL method behaves
in the same way as QCD. However, we shall see later that the quantitative
predictions are different.

\section{ Weak field.}

We will consider QCD with two massless flavors.
In the leading order of chiral perturbation theory, the
masses of $u-$ and $d-$ quarks appear in the vacuum energy only 
 via its dependence on $M_{\pi}^2$ which is proportional to the sum $m_u +m_d$.
It means that in this order $<\bar{u} u>$ and $<\bar{d} d>$
 condensates
will not differ and we can define the quark condensate as 
\be
\Sigma(H)\ = \ - \frac{\partial \epsilon_{vac} (m_u,m_d,H)}
{\partial m_u}\left|_{m_u=m_d=0}\ = \
 - \frac{\partial \epsilon_{vac} (m_u,m_d,H)}{\partial m_d}\right|_{m_u=m_d=0}
\label{Fcond}
\ee
where the small quark masses $m_u, m_d $ are introduced as  
external probes.

 To find $\epsilon_{vac}$, we need to calculate vacuum
loops in the presence of magnetic field. When the field is weak
$eH\ll \mu_{hadr}^2 \sim {(2\pi F_\pi)}^2$ (however, we will always assume
$eH\gg {M_\pi}^2$), characteristic momenta in the loops are small and
the theory is adequately described by the effective low energy chiral
Lagrangian  ($\ref{CLLO}$). 
In the leading order, pion interactions can be neglected
whatsoever, and the field dependent part in $\epsilon_{vac}$ is given by
one loop graphs depicted in Fig.~1.

\begin{figure}
\centerline{\epsfbox{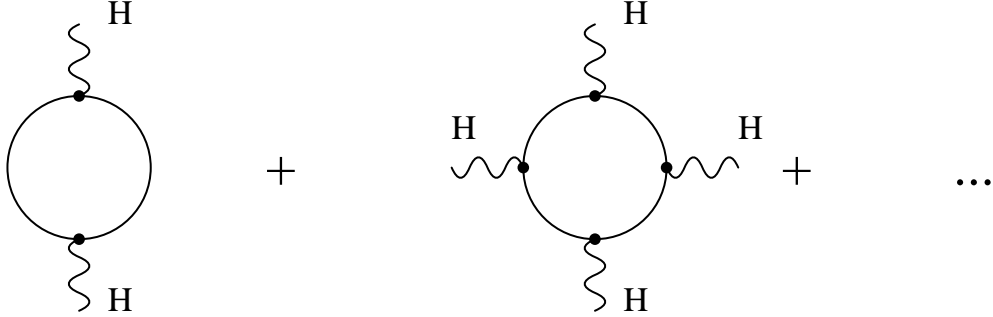}}
\caption{ Vacuum energy in a weak magnetic field. Solid lines
stand for charged pions.}
\end{figure}

 It is instructive to consider first the graph with two photon legs.
 The corresponding contribution in the vacuum energy is 
 \be
 \epsilon^{(2)}_{vac}(H, {M_\pi}^2)=\frac{e^2 H^2}{96\pi^2}
 \ln\frac{\Lambda^2}{{M_\pi}^2},
\label{TPF}
 \ee
 where $\Lambda$ is the ultraviolet cutoff. By virtue of 
 Gell-Mann - Oakes - Renner relation,
 \be
{F_\pi}^2 {M_\pi}^2 = \Sigma ( m_u + m_d),
 \label{GOR}
 \ee
one can relate the derivatives over $m_u (m_d)$ and over ${M_\pi}^2$.
Taking into account Eqs. ($\ref{Fcond}$), ($\ref{TPF}$),
and ($\ref{GOR}$), we obtain
 \be
 \Delta \Sigma ^{(2)}(H)=\frac{\Sigma}{{F_\pi}^2}
 \frac {e^2 H^2}{96\pi^2 {M_\pi}^2}.
\label{TPR}
 \ee
 This expression diverges in the chiral limit ${M_\pi}^2 \rightarrow 0$
 and makes as such little sense. It is easy to trace back the origin
 of this divergence. The corresponding graph for $\epsilon_{vac}$
 involves the logarithmic divergence both in the ultraviolet and in the infrared.
 After differentiating, it gives the contribution 
 in the condensate which diverges as a power in infrared.

 But one is not allowed, of course, to restrict oneself by the graphs
 with only two external field insertions. All other graphs are also
 important. The more is the number of legs, the more severe are the 
 infrared singularities. For example, the graph with four legs involves
 the singularity $\sim (eH)^4/ {M_\pi}^4$ in the vacuum energy which
 gives the singularity $\propto 1/{M_\pi}^6$ in the condensate, etc.
 All such graphs should be summed up. The result is the analog of
 the Euler-Heisenberg Lagrangian for scalar particles which was found
 long time ago by Schwinger \cite{Schw}. We have 
\footnote{This is the unrenormalized vacuum energy. More 
precisely, the quartic divergence which corresponds to the loop without
legs and is field independent is subtracted, but the logarithmic
ultraviolet divergence associated with the graph with two legs is not.
The expression usually found in the textbooks corresponds to subtracting
also the two-leg graph, so that the expansion of \ 
$\epsilon_{\rm vac}^{\rm ren}(H,M_\pi)$ in $H$ starts from the term 
 $\propto H^4$. The two-leg graph is then absorbed in the charge
( $\equiv$ field ) renormalization, and the derivative of 
$H^2_{\rm ren} (M_\pi)/2 + \epsilon_{\rm vac}^{\rm ren} (H,M_\pi)$  over mass
would, of course, be the same as that of $\epsilon_{\rm vac}^{\rm unren}$.}
 \be
 \epsilon_{vac}=-\frac{1}{16\pi^2} \int_0^\infty\frac{ds}{s^3}
 e^{-{M_\pi}^2 s} \left[\frac {eHs}{\sinh(eHs)}-1\right].
\label{Sch}
 \ee
 We see that the integral ($\ref{Sch}$) is regular in the infrared 
 (large s region). The infrared cutoff is provided now not by 
 ${M_\pi}^2$, but by the field $eH$ itself. All infrared divergent
 pieces in the graphs in Fig. 1 are summed up into a finite expression.

 To find the shift of the condensate, we have to differentiate 
Eq. ($\ref{Sch}$)
 over quark mass and substitute it into the 
 definition ($\ref{Fcond}$). We arrive at the expression proportional
to the simple integral 
\be
I \ = \ \int^\infty_0 \frac{dz}{z^2}\left\{\frac{z}{\sinh(z)}-1 \right\} 
\ =\ -\ln 2
\label{Int}
\ee
which can be done as
 a sum of residues of the poles on, say, positive imaginary $z$ axis.
 Our final result is 
 \be
 \Sigma(H)=\Sigma(0)\left[1+\frac{eH \ln2}{16\pi^2 {F_\pi}^2}+...
\right].
\label{Res}
 \ee
 We see that the shift is positive and linear in $H$. The latter is easy
 to understand if substituting the actual infrared cutoff $\sim eH$ for 
$M_{\pi}^2$ in Eq. ($\ref{TPR}$).

It is also clear now why NJL model gave the shift which was quadratic
rather than  linear in field: the corresponding calculation involved the loop
of massive quarks rather than the loop of massless pions and was infrared
finite in any order in $H$. A side remark is that a simple-minded NJL
calculation is also not able to reproduce the result ($\ref{conT}$) for 
the temperature 
dependence of the condensate. At small temperatures, the density of
massive quarks in the heat bath is exponentially suppressed, while massless
pions are in abundance.

The correction $\propto eH$ in Eq. ($\ref{Res}$) is of order 1
 when the field is 
comparatively large $\sqrt {eH} \sim 1.4$ GeV. Well before that, our
calculation based on the effective pion Lagrangian ($\ref{CLLO}$) 
loses validity. In 
principle, one can trace the deviation from the leading order result 
($\ref{Res}$) at
intermediate values of $H$ in the framework of chiral perturbation theory
\cite{CPT} taking into account nonlinear pion interactions. 
An example
of the graph contributing in the vacuum energy in the next order is depicted in
Fig. 2. Also higher order terms in the chiral Lagrangian which we did not
specify and did not discuss here would become important. The estimate
for two-loop correction to the condensate is $\propto \Sigma(0){(eH)^2}/
{(2\pi {F_\pi})}^4 $ (the same estimate $\propto \Sigma(0){(eH)^2}/
\mu_{hadr}^4 $ is obtained if taking into account massive charged
particles such as $K$  and $\rho$ meson). In this order, one should not
expect the corrections to $|<\bar u u>|$ and $|<\bar d d>|$ to be the same
(the magnetic field breaks down isotopic invariance). The calculation of the 
coefficients
(it is not yet clear whether also some logarithmic factor $\propto 
\ln\frac{eH}{\mu_{hadr}}$ appears ) is now in progress.

\begin{figure}
\centerline{\epsfbox{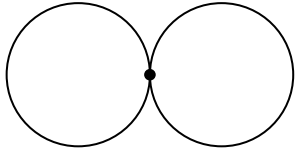}}
\caption{ A two-loop graph contributing to the
vacuum energy.}
\end{figure}

\section{$M_{\pi}(H)$, $F_\pi (H)$ and Gell-Mann - Oakes - Renner 
relation.}

As the electric charges of $u-$ and $d-$ quarks are different, flavor 
symmetry is broken in an external magnetic field. In particular, the axial
$SU_A (2)$ symmetry is broken down to $U_{A}^3 (1)$ corresponding
to chiral rotation of $u-$ and $d-$ quarks  with opposite phases ( the
singlet axial symmetry is broken already in the absence of the field due
to the anomaly ). The formation of the condensate breaks down this remnant
$U_{A}^3 (1)$ symmetry spontaneously leading to appearance of a 
Goldstone boson, the $\pi^0$-meson. Indeed, charged pions acquire 
a gap in the spectrum $\propto \sqrt {eH}$ and are not goldstones
 anymore.   If quarks are endowed a small non-zero mass $m$, pions
are not exactly massless, their mass being related to the quark condensate
by the  Gell-Mann - Oakes - Renner relation ($\ref{GOR}$).

It is interesting to study the question how the mass and residue of $\pi^0$
depend on the external field.

Let us find first the mass shift. We have to calculate the polarization operator
of $\pi^0$ given by the sum of graphs of the kind drawn in Fig. 3.

\begin{figure}
\centerline{\epsfbox{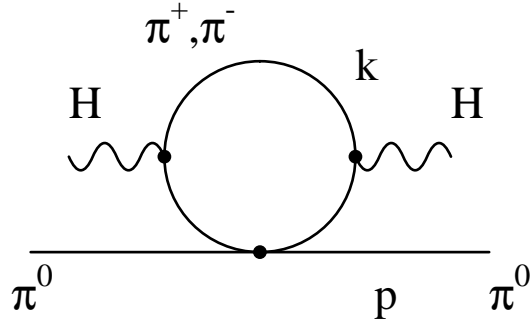}}
 \caption{ An example of the graph contributing to the polarization operator 
of $\pi^0$ in a magnetic field.}
\end{figure}

The four-pion vertex can be found from the expansion of the effective chiral
Lagrangian. In the exponential parameterization 
$U= \exp\{i \phi^a \tau^a/F_\pi \}$,
the corresponding terms in the Lagrangian have the form
\be
L^{(4)}=\frac{1}{6 F_{\pi}^2} [{(\phi^a\partial_\mu \phi^a) }^2-
(\phi^a\phi^a){(\partial_\mu \phi^b)}^2]+
\frac{M_{\pi}^2}{24F_{\pi}^2}{(\phi^a\phi^a)}^2.
\label{L4}
\ee

We need also the expression of charged pion propagator in a magnetic
field. It can be inferred from the results of Ref.  \cite{Chodos} 
where an explicit expression
for the fermion propagator in a magnetic field at non-zero chemical potential
$\mu$ has been found. Basically, one has to
take the integral multiplying the factor $-\mu\gamma^0$ in Eq. (4.9) of 
Ref. \cite{Chodos}
 without the factor $\cos(eHs)+
\gamma^1 \gamma^2 \sin(eHs)\equiv \exp\{ie(\sigma F)s/2\}$ in the 
integrand.  We arrive at the following expression for the Euclidean scalar
propagator
\be
D_{H} (x,y)= \exp\left\{ie\int^x_y A_\mu(\xi) d\xi_\mu \right\}
\int \frac{d^4 k}{{(2\pi)}^4}e^{-ik(x-y)} D_{H} (k),
\label{DHxy}
\ee
where the integral in the phase factor is done along the straight line connecting
$x$ and $y$, and 
\be
D_H (k)=\int_0^\infty \frac{ds}{\cosh(eHs)}
 \exp\left\{-s\left[k_{\parallel}^2 + k_{\perp}^2\frac{\tanh(eHs)}{eHs}+
M_{\pi}^2\right]\right\}
\label{DHk}
\ee
with $k_{\parallel}^2=k_{4}^2+k_{3}^2 $,
 $k_{\perp}^2=k_{1}^2+k_{2}^2$
(the magnetic field is aligned along the third axis).

Substituting ($\ref{DHk}$) and the vertex inferred from ($\ref{L4}$) 
in the graph in Fig. 3 (as the
propagator enters with coinciding initial and final points, the phase factor
disappears), and subtracting the similar expression at $H=0$, we obtain
$$
\Pi_H(p^2)-\Pi_H (0)=-\frac{1}{3 F_{\pi}^2}
\int \frac{d^4 k}{{(2\pi)}^4}
(M_{\pi}^2+2p^2+2k^2)
\int_0^\infty ds \left[\frac{ \exp\{-s(k_{\parallel}^2 +
 k_{\perp}^2\frac{\tanh(eHs)}{eHs}+M_{\pi}^2)\}}{\cosh(eHs)} \right.
$$
\be
\left.
-\exp\{-s(k_{\parallel}^2+k_{\perp}^2+M_{\pi}^2)\}\right].
\label{PiH}
\ee

The mass shift is given by the shift ($\ref{PiH}$)
 of the (Euclidean) polarization
operator at the point $p^2=-M_{\pi}^2$. Note first of all that the shift is 
zero for massless pions. An exact goldstone remains the exact goldstone
also when $H\neq 0$.  The shift $\Delta_H M_{\pi}^2$ is proportional to
$M_{\pi}^2 (H=0)$. Taking two first terms of the expansion of 
$M_{\pi}^2$ and calculating first the integral over 
$d^4 k=\pi^2 d k_{\parallel}^2 d k_{\perp}^2$ and then over $ds$ 
(the latter has the same structure as the integral ($\ref{Int}$) 
which we encountered when
calculating the shift of the condensate and involves also an extra piece 
$\sim \int_0^\infty ds[\cosh(s)/\sinh^2(s) - 1/s^2]$ which is, however,
zero), we arrive at the simple result 
\be
\Pi_H(p^2)-\Pi_0 (p^2) =-p^2\frac{eH \ln2}{24\pi^2 F_{\pi}^2}+
M_{\pi}^2 \frac{eH \ln2}{48\pi^2 F_{\pi}^2}+...
\label{PiHres}
\ee
which implies
\be
M_{\pi}^2 (H)=M_{\pi}^2 (0)\left[1-\frac{eH \ln2}{16\pi^2 F_{\pi}^2}+...
\right]
\label{MH}
\ee

\begin{figure}
\centerline{\epsfbox{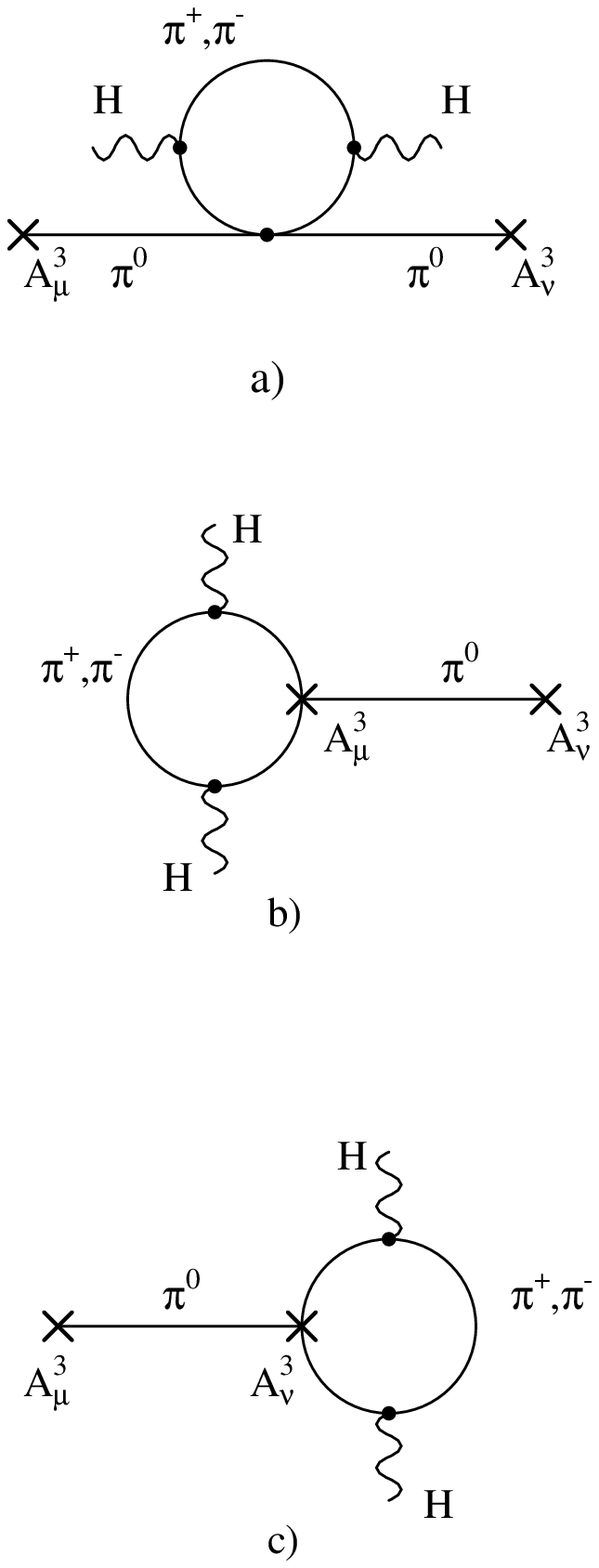}}
 \caption{Axial currents correlator in a magnetic field.
The graphs contain exact propagator of charged pions in
the constant magnetic field.}
\end{figure}

Let us find now the renormalization of the residue brought about by the field.
We need to calculate the one-loop graphs depicted in Fig. 4  which 
contribute to the pole structure  in the axial current correlator
$\int dx e^{ipx} <A_{\mu}^3 (x)  A_{\nu}^3  (0)>_H d^4 x \sim
p_\mu p_\nu F_{\pi}^2 (H)/(p^2+M_{\pi}^2)$.
As $F_\pi$ and its shift are non-zero in the chiral limit, we can set 
$M_{\pi}^2=0$. The graph in Fig. 4a has already been calculated
(the first term in Eq. ($\ref{PiHres}$) ). 
To calculate the graphs in Fig. 4b,c , we need the
vertex $<0|A_{\mu}^3 |\pi^0 \pi^+ \pi^- >$ which can be found by
"covariantizing" the derivative in Eq. ($\ref{CLLO}$) 
according to the rule \cite{CPT}
$\partial_\mu U \rightarrow \partial_\mu U -i(A_\mu U+U A_\mu)$
where $A_\mu=A_\mu t^a$. The final result is
\be
F_{\pi}^2 (H)=F_{\pi}^2 (0) \left[1+\frac{eH \ln2}{8\pi^2 F_{\pi}^2}
+...\right]
\label{FH}
\ee
Note that the contributions of the individual graphs Fig. 4a and Fig. 4b,c 
depend on the chosen exponential parameterization, but the physical pion
residue $F_\pi(H)$ depending on their sum does not. Also the physical pole
 position ($\ref{MH}$)
does not depend on parametrization while the individual terms in 
($\ref{PiHres}$) do.

The results ($\ref{Res}$), ($\ref{MH}$), and  ($\ref{FH}$)
 are very much analogous to Eq. ($\ref{conT}$) and the similar known
expressions for $M_{\pi}^2 (T)$ and $F_{\pi}^2 (T)$. Only here the 
physical situation is just opposite compared to the thermal case. A nonzero
temperature suppresses the condensate and residue, and leads to the 
increase of the mass. A nonzero magnetic field brings about the increase
in the condensate and residue, and suppresses the mass.

In the thermal case, the  renormalized condensate, pion mass, and residue
still satisfy the relation ($\ref{GOR}$) in the leading order in $T^2$.
 Likewise, the 
expressions ($\ref{Res}$) ,($\ref{MH}$), and ($\ref{FH}$)
 satisfy the  Gell-Mann -- Oakes -- Renner 
relation in the leading order in $eH$.

\section{Strong field.}

When $eH\gg \mu_{hadr}^2$, characteristic momenta in the vacuum 
loops are high, and the system is adequately described in terms of 
quarks and gluons rather than in terms of pions and other low lying
hadron states. Due to asymptotic freedom, the effective coupling
constant $\alpha_s (eH)$ is small, and we can try to treat strong 
interaction effects pertubatively.

For sure, in QCD with massless quarks, the condensate cannot 
appear in any finite order of perturbation theory, irrespectively of
whether a magnetic field is present or not --- both electromagnetic and
strong interaction vertices respect chirality.

It was recently discovered, however, that the condensate is 
generated in the strong field limit \cite{Kiev}. To see that, one has 
to sum up an infinite set of relevant graphs. The authors of \cite{Kiev}
 studied the Bethe-Salpeter equation which implements such a resummation  
for massless
QED and showed that the equation admits a nontrivial solution
with dynamically generated mass. Later this result was reproduced
in the language of Schwinger-Dyson equation \cite{Ng}. The derivation applies 
also
in QCD without essential modifications. For clarity sake, let us say
here  few words about it.

When strong interaction is disregarded whatsoever, we have a  
system of free charged quarks in a magnetic field. The spectrum
presents the set of Landau levels
\be
\epsilon_{\pm}(n,\sigma ,k_3 )=\pm\sqrt{|e_q H|(2n+\sigma +1)+
k_{3}^2+m_{q}^2},
\label{spectr}
\ee
where $n=0,1,...$,\ $\sigma=\pm 1$ marks the spin orientation, and
$k_3$ is the momentum along the magnetic field direction. 
Negative energies describe the Dirac sea. The corresponding
eigenstates are localized in the transverse direction. The 
characteric size of the orbits is $\sim 1/\sqrt{eH}$ ( so that for 
$\sqrt {eH}\sim 1.4$ GeV they are already pretty small). In infinite
space, each state ($\ref{spectr}$)
 is infinitely degenerate, the degeneracy being
associated with the position of the center of the orbit. For a finite box
of size $L$, the level of degeneracy is
\be
N_\perp=\frac{|e_q H|}{2\pi} L^2,
\label{LDe}
\ee
and also $k_3$ is quantized to $2\pi n/L$. For $m_q=0$, the 
ground states have zero energy.

We need in the following the fermion Green's function in 
a magnetic field. Its explicit expression was found in \cite{Skob1}
and, in a somewhat more convenient form, in \cite{Chodos}
 with Schwinger 
technique. First of all, one can write
\be
G_H (x,y) = \exp\left\{i e_q \int^x_y A_\mu (\xi) d\xi_\mu \right\} 
\hat{G}_H (x-y)
\label{fGH}
\ee
where the integral in the phase factor is done along the straight 
line connecting $x$ and $y$. The Fourier image of 
$\hat{G}_H (x-y)$ presents a complicated integral over proper 
time. Fortunately, we do not need the full expression, but only
its asymptotic form in the region of small momenta $k\ll \sqrt {eH}$.
In this region, it suffices to retain only the lowest landau  levels 
(LLL) with $n=0$, $\sigma=-1$ in the spectral decomposition of 
the Green's function, and the latter acquires the simple form 
\cite{Skob1,Kiev}
\be
\hat {G}_H (k)=ie^{(-k_{\perp}^2/|e_q H|)}\quad
\frac{\hat{k}_{\parallel}+m_q}{k_{\parallel}^2-m_{q}^2} 
(1-i\gamma^1 \gamma^2),
\label{Gr}
\ee
where $k_{\parallel}^2=k_{0}^2-k_{3}^2$ and 
$k_{\perp}^2=k_{1}^2+k_{2}^2$.

Basically,  this Green's function describes a free motion of the 
states with $\sigma=-1$ in longitudinal direction. Strictly 
speaking, retaining the exponential factor 
$ \exp(-k_{\perp}^2/{|e_q H|})$ is not quite consistent - 
the dominance of LLL on which the derivation of Eq. ($\ref{Gr}$)
 was 
based is justified only in the region where both $k_{\perp}^2$ and
$k_{\parallel}^2$ are small compared to $|e_q H|$ and the 
exponential factor is not efficient. We will see later, however, that the
generation of  condensate is related to the infrared region 
$k\ll\sqrt {|e_q H|}$. Really, in the opposite limit  $k\gg\sqrt {|e_q H|}$,
the  Green's function $G_H (x,y)$ tends to the free fermion Green's
function, and we cannot expect a nontrivial dynamic phenomenon
like the condensate generation to be associated with that region. The 
exponential factor in ($\ref{Gr}$) is convenient to retain as an 
effective 
momentum cutoff.

We are interested in the theory with massless quarks, so that 
$m_q=0$ in the first place. We expect, however, the dynamical mass
generation, and our Ansatz for the exact Green's function at low 
momenta is Eq. ($\ref{Gr}$) with nonzero and, generally speaking,
momentum-dependent $m_q$.

Next, we substitute this Ansatz in the Schwinger-Dyson equation schematically
presented in the Fig. 5. The equation was solved in the approximation where
the loop corrections to the vertices and also to the gluon propagator (the latter
is actually a rather strong and not so innocent assumption) were disregarded.

\begin{figure}
\centerline{\epsfbox{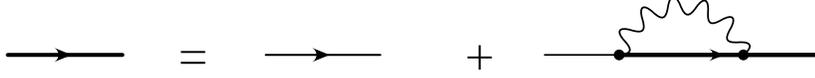}}
 \caption{Schwinger-Dyson equation for the exact quark
propagator. Thin solid lines correspond to the free 
Green's function and bold lines - to the exact one. }
\end{figure}

Doing the spinor trace and performing Wick rotation, one can see after
some transformations that the equation admits a factorized
solution
\be
m_q ({\bf p_\parallel} , {\bf p_\perp})=\hat{m}_q
({ p_\parallel}^2)
\exp\{-{ p_{\perp}^2}/|e_q H|\},
\label{mp}
\ee
where $\hat{m}_q ({ p_\parallel}^2)$ satisfies the following 
2-dimensional integral equation 
\be
\hat{m}_q ({ p_\parallel}^2)=\frac{\alpha_s c_F}{2\pi^2}\int
\frac{d^2 {k_\parallel}\  \hat{m}_q ({ k_\parallel}^2) } 
{{ k_{\parallel}^2}+\hat{m}_{q}^2({ k_\parallel}^2)} \int
\frac{d^2 { k_\perp} \exp\{-{{ k_{\perp}^2}}/(2|e_q H|)\}}
{{({\bf k_\parallel} -{\bf p_\parallel})}^2+{ k_{\perp}^2}}.
\label{meq}
\ee
The integral ($\ref{meq}$) is double logarithmic
\footnote{The double logarithmic structure of the integral is
easy to understand if disregarding the phase factor in 
Eq. ($\ref{fGH}$)
which makes the calculations elementary. That gives 
qualitatively the same form of the integral as in 
Eq. ($\ref{meq}$).
Quantitative results, however, would be wrong. In particular,
the coefficient in Eq. ($\ref{meqf}$) would be 4 times larger 
than the actual one.}
 receiving a main support from
the region $\hat{m_q}^2 (0)\ll k_{\parallel}^2\ll k_{\perp}^2
\ll |e_q H|$

Let us first tentatively neglect the dependence of 
$\hat{m_q}$ on longitudinal momenta. We would obtain
\be
\hat{m_q} (0) \sim \hat{m_q} (0)\frac{\alpha_s c_F}{4\pi}
\ln^2 \frac {|e_q H|}{\hat{m_{q}^2} (0)}.
\label{meqf}
\ee
The corresponding calculation was actually first done in \cite{Janc}
where a {\it correction} to the electron mass in the presence of the
external magnetic field $\Delta m(H) = 
\frac{\alpha m(0)}{4\pi} \ln^2 (eH/m^2(0))$ was found. In that (and other)
earlier papers, the equation (\ref{meqf}) was not treated, however,
as a self-consistent equation  allowing to unravel the dynamical
generation of mass even if the lagrangian electron mass is zero.
The solution of this equation is 
\be
\hat{m_q} (0)\sim \sqrt {|e_q H|} 
\exp\left\{-\sqrt {\frac{\pi}{\alpha_s c_F}}\right\}
\ee
The exponential factor displays a truly nonpertubative 
nature of the result. An accurate analysis of Ref. \cite{Kiev}
 which takes
into account the  momentum dependence of the mass leads to the result
\be
\hat {m} \sim \sqrt {|e_q H|} 
\exp \left\{-\frac{\pi}{2}\sqrt{ \frac{\pi}{2\alpha_s c_F}}\right\}
\ee
The quark condensate is defined as 
\be
<\bar{q}q>_H=-\int \frac{d^4 p}{{(2\pi)}^4}\ Tr\hat{G}_H (p)=
-\frac{4}{{(2\pi)}^4}\int d^2 p_\perp d^2 p_\parallel\ 
\frac{m_q (p_{\parallel}^2, p_{\perp}^2)}
{p_{\perp}^2+p_{\parallel}^2+m_{q}^2 
(p_{\parallel}^2, p_{\perp}^2)}
\ee
Assuming the exponential fall-off of dynamical mass at
$p_{\perp}^2\gg |e_q H|$, $p_{\parallel}^2\gg |e_q H|$
(cf. Eq. ($\ref{mp}$) ), we finally obtain
\be
\Sigma_H=-<\bar{q}q>_H\  = \ F(\alpha_s){|e_q H|}^{3/2}
\exp\left\{-\frac{\pi}{2} 
\sqrt{\frac {\pi}{2\alpha_s (|e_q H|) c_F}}\right\}
\label{FCLH}
\ee
where $F(\alpha_s)$ is some function not involving an
exponential dependence.

We see that the condensate increases with the field, and no
phase transition occurs.

\section{Discussion.}

Our main result ($\ref{Res}$) is an exact theorem of QCD. 
It has the same
status as Eq. ($\ref{conT}$) and some exact
 results for density of 
eigenvalues of Euclidean Dirac operator at finite volume 
\cite{LS}
and in thermodynamic limit \cite{Stern}. 
It would be very interesting
to check this and other similar exact theorems in lattice 
calculations with dynamical fermions and/or vacuum model 
calculations
\footnote{Unfortunately, lattice calculations are very difficult
here. The Euclidean lattice should be roomy enough to
accommodate pions: the size of the box should be considerably
larger than the pion Compton wavelength. Calculations in the 
framework of the instanton liquid model \cite{Shur}
 allow for larger 
boxes and are more promising.}.

The derivation of the result ($\ref{FCLH}$) in the strong 
field limit
reviewed in the previous section is similar in spirit to the
derivation of spontaneous symmetry breaking in the NJL model.
In both cases, a self-consistent gap equation for the fermion
propagator is solved. This method of derivation seems to be 
rather reliable
\footnote{There is, however, a not quite clear for us at the moment
question on whether the loop corrections to the gauge boson line
in the Schwinger -- Dyson equation in Fig.5  could modify the result. According
to  \cite{Skob2}, the effective mass
which gauge bosons acquire  in a magnetic field
brings about an additional infrared cutoff which kills one
of the logarithms in the strong field asymptotic
of the fermion mass operator }
, but it is indirect. It would be very nice to 
understand the appearance of the quark condensate in strong
fields in plain and direct terms.

We have in mind the following. According to the famous Banks
and Casher theorem \cite{Banks}, 
the quark condensate is related to the
spectral density of Euclidean massless Dirac operator 
$\rho (\lambda )$ at $\lambda =0$
\be
\Sigma =\pi \rho (0).
\ee
Nonzero spectral density $\rho (0)$ means that the number
of Dirac operator eigenvalues with 
$\lambda \leq\lambda_0 \ll \mu_{hadr}$ is estimated as 
$N(\lambda \leq \lambda_0 )=\rho (0)\lambda_0 V,$ where $V$
is the Euclidean volume. For free massless fermions, the spectrum is
$\lambda_{\bf n}=2\pi |{\bf n}|/L$ where ${\bf n}$ is a four- 
dimensional integer vector. $\rho_{free}(\lambda) \sim
\lambda^3,$ $\rho_{free}(0)=0$ and the condensate is zero.
For free massless fermions in a magnetic field, the infrared branch of
the spectrum is 
\be
\lambda_{\bf m}=\frac{2\pi}{L} |\bf {m}|,
\label{fre}
\ee
where $\bf m$ is a two-dimensional integer vector. The 
eigenvalues ($\ref{fre}$) correspond to free motion
 in the longitudinal
direction. Each level ($\ref{fre}$) involves the high level of 
degeneracy ($\ref{LDe}$). 
The spectral density is then estimated as 
$\rho_{free}^H (\lambda )\sim |e_q H|\lambda$. 
This is "better" than
in the absence of the field, but still $\Sigma=\pi\rho(0)=0$.
A nonzero condensate ($\ref{FCLH}$) means that, 
after taking into account
the interaction between quarks, the Euclidean Dirac spectrum
($\ref{fre}$) is substantially modified, 
and small eigenvalues with
characteristic spacing $\sim 1/(\sqrt {|e_q H|}L^2)$ 
appear. We are not able now to display the presence of such 
small eigenvalues explicitly.

An external magnetic field increases the condensate which means 
that it should make the chiral restoration phase transition in
temperature and/or in baryon chemical potential more
difficult. That means, in particular, that critical 
temperature $T_c$ (at $H=0$, it is estimated to be of order
$200$ MeV \cite{review}) should increase with $H$. 
This question was 
studied in recent \cite{Ng1} for massless QED. 
Translating their
result in QCD language, we obtain 
\be
T_c \sim \alpha_s \sqrt{|e_q H|}
\label{THres}
\ee
in the strong field region. The result ($\ref{THres}$)
 is easy to 
understand. $\sqrt{|e_q H|}$ appears by dimensional reasons
and the factor $\alpha_s$ appears because the generation of
the condensate in a strong magnetic field is driven by strong perturbative
inter-quark interaction.

\section{\bf Acknowledgements}

One of the authors (A.V.S.) acknowledges an illuminating
discussion with A.A.Migdal which precipated his interest
in this problem. We are indebted to B.O. Kerbikov who
pointed our attention to the paper \cite{Klev} 
and to B.L. Ioffe,  A.S. Vshivtsev,  and V.V. Skobelev for
very useful critical discussions. This work  was done under
the partial support of INTAS grants 93-0283 and 94-2851, CRDF grant RP2-132,
 and Schweizerischer National Fonds grant 7SUPJ048716.
A.V.S. acknowledges  warm hospitality extended to him at Univ. of Minnesota.

\end{document}